\documentclass[
 reprint,
 superscriptaddress,
 amsmath,
 amssymb,
 aps,
 prb
]{revtex4-2}

\usepackage[colorlinks=true,allcolors=blue]{hyperref}
\usepackage{graphicx}
\usepackage{dcolumn}
\usepackage{bm}

\usepackage[dvipsnames]{xcolor} 

\begin{document}

\title{Fully screened two-dimensional magnetoplasmons and rotational gravity shallow water waves in a rectangle}

\author{D.A. Rodionov}%
\email{rodionov.da@phystech.edu}
\affiliation{Kotelnikov Institute of Radioengineering and Electronics of Russian Academy of Sciences, Moscow, 125009 Russia}
\affiliation{Moscow Institute of Physics and Technology, Dolgoprudny, Moscow Region, 141701 Russia}

\author{I.V. Zagorodnev}%
\affiliation{Kotelnikov Institute of Radioengineering and Electronics of Russian Academy of Sciences, Moscow, 125009 Russia}

\date{\today}

\begin{abstract}

We study plasmons in a rectangular two-dimensional (2D) electron system in the vicinity of a planar metal electrode (gate) and in the presence of a perpendicular uniform magnetic field, using Maxwell’s equations and neglecting retardation effects. The conductivity of the 2D system is characterized by the dynamical Drude model without taking collisional relaxation into account, which well describes both high mobility graphene and other field effect transistor structures, including quantum wells like Ga(Al)As, in the terahertz and in some cases sub-terahertz frequency ranges. Without a magnetic field, we analytically find the current distribution and frequency of plasma eigenmodes when the plasmon wavelength is much larger than the distance to the gate, i.e. in the fully screened limit. To find an approximate solution in a magnetic field, we expand current in the complete set of eigenmodes without magnetic fields. Analytical asymptotic expressions in weak and strong magnetic fields were obtained for the lower modes. Unlike the disk and stripe, the frequencies of these modes tend to zero as the magnetic field tends to infinity. We also discuss a direct analogy to rotational gravity shallow water waves, where size-quantized Poincare waves correspond to size-quantized magnetoplasmons, while Kelvin waves correspond to edge magnetoplasmons.

\end{abstract}

\maketitle

\section{\label{sec:introduction} Introduction}

The collective electron oscillations in 2D electron systems (ESs), called 2D plasma waves or 2D plasmons, have been actively studied since 1967 \cite{Stern1967,Grimes1976,Allen1977,Theis1977}, in particular, in connection with sub-terahertz and terahertz detection \cite{Knap2009,Muravev2012,Bandurin2018,Caridad2024}. For a long time they have been scrutinized in structures like metal-oxide-semiconductor field-effect transistors or high-electron-mobility transistors \cite{Tsui1980,Ando1982,Kukushkin2003,Lusakowski2016}, however, recently, they have been intently discussed in graphene and other novel 2D materials \cite{Grigorenko2012,Yao2018,Mitin2020}, where they are also called surface plasmon polaritons.

When a system consists only of a 2D conducting layer immersed in a homogeneous dielectric medium (``unscreened'' or ``ungated'' system), the frequency of 2D plasmons has a square root dependence on the wavevector as well as on the electron concentration in the simplest quasielectrostatic approximation. The electron concentration and therefore the frequency of 2D plasmons can be controlled by applying voltage to a planar metal electrode (gate) placed near 2D ES due to the field effect. The gate also screens the electric fields of charges and therefore, even without any voltage on it, changes the long wavelength dispersion of the plasmons to a linear one \cite{Chaplik1972,Ando1982}. The latter takes place when the plasmon wavelength is much larger than the distance to the gate and this is sometimes called the fully screened limit \cite{Fetter1986,Zagorodnev2022} or the local capacitance approximation \cite{Volkov1988} since the charge density is proportional to the electric potential locally, as for a parallel plate capacitor. In this limit, the theory of plasmons is significantly simplified, and an analytical description of a number of laterally confined 2D ESs, e.g., disk and stripes, was obtained \cite{Fetter1986,Svintsov2018,Volkov1988,Zagorodnev2023,Rodionov2023}.

The dispersion law of 2D plasmons also depends on the external magnetic field. Surprisingly, even within the fully screened limit, the analytical description of 2D plasma spectra in a square in a perpendicular uniform magnetic field remains still an unsolved problem. Nevertheless, recently, magnetoplasmons in an unscreened square-shaped 2D ES based on GaAs quantum well have been investigated experimentally \cite{Zarezin2023}. It turned out that a hybridization of neighboring magnetoplasmon modes leads to anticrossing of magnetodispersions of some modes (avoided crossings), whereas no such hybridization was observed in the disk-shaped geometry. Inspired by the experiment and numerical calculations \cite{Mashinsky2023}, we analyze the plasmons in a rectangular 2D ES, in particular in a square, in the fully screened limit. We describe plasmons in the simplest classical manner using Poisson's equation, continuity equation, and local Ohm's law with the dynamical Drude conductivity, assuming that the plasmon frequency is much less than the Fermi energy divided by $\hbar$, the interband transition frequencies, and the wavelength of the plasmon is much greater than the Fermi wavelength or the electron mean free path. Unlike Ref. \cite{Zarezin2023}, we use a complete basis set of functions, which is merely the solutions in zero magnetic field, to obtain accuracy-controllable approximation for plasma frequencies, including high magnetic fields.

Before further discussing magnetoplasmons, we would like to draw attention to a very similar hydrodynamic system. Thus, in Ref. \cite{Dyakonov1993} it was noted that in the fully screened limit the electron dynamics in a 2D ES (without a magnetic field) is similar to shallow water dynamics when the time of electron-electron collisions is less than the electron relaxation time due to collisions with impurities and phonons, i.e., at intermediate, somewhere between room and helium, temperatures \cite{Zabolotnykh2023}. This analogy was also extended to equatorial waves on a rotating sphere (planet) \cite{Finnigan2022}. The role of the Coriolis force is played by the Lorentz force due to the external magnetic field. We explicitly show in Sec. II that rotational gravity waves in a pool, also known as Poincare waves \cite{Vallis2017}, exactly correspond to the fully screened magnetoplasmons in the same shape structure. In particular, the Kelvin waves existing near the linear edge of a pool correspond to the edge magnetoplasmons. Therefore, by characterizing magnetoplasmons in a rectangular fully screened 2D ES, we simultaneously describe rotational gravity waves in shallow water. As far as we know, no analytical solution in a square has been found for the latter. Besides, we clarify that the fully screened plasmon shallow water analogy is also correct when the 2D ES is described by the dynamical (or ``optical'') Drude conductivity without taking collisions into account. In particular, the Drude conductivity describes well 2D ESs for low (``helium'') temperatures in the gigahertz and terahertz frequency ranges \cite{Burke2000,Cui2021,Shuvaev2021}. Thus, the analogy discussed above is actually stronger and can be applied to plasmons not only at intermediate temperatures but also at low temperatures, for which the time of electron-electron collisions is greater than the electron relaxation time.

\section{\label{sec:main_body} Magnetoplasmon - shallow water analogy}

Let us consider a 2D ES of arbitrary geometric shape, placed in the plane $z=0$ on a substrate with thickness $d$ and dielectric permittivity $\varepsilon$. The infinite gate is located in the $z=-d$ plane. An external uniform magnetic field is perpendicular to the 2D ES. We assume that the current density, the perturbation of the charge density, and the electromagnetic fields oscillate at frequency $\omega$, i.e. they are proportional to $e^{-i\omega t}$. We examine plasmons in the quasistatic limit, when the light velocity $c$ in Maxwell's equations formally approaches to infinity. In this case, magnetic fields are negligible, and electric fields are completely determined by the scalar electric potential $\varphi(x,y,z)$. Considering that the distance $d$ is much smaller than all the lateral sizes of the 2D ES and plasmon wavelength, Poisson's equation is reduced to the following local connection between the electric potential $\varphi(x,y,z,t)$ and the charge density $\delta\rho(x,y)$:
\begin{equation}\label{eq:potential}
    \varphi(x,y,z=0)=\frac{4\pi d}{\varepsilon} \delta\rho(x,y).
\end{equation}
A detailed derivation of the equation is presented in Appendix \ref{app:A}. Here and below we use the CGS system. This fully screened limit in fact corresponds to the local interaction: electrons in the 2D ES interact mainly with their image charges in the metal gate. This limit can be easily realized in real samples \cite{Bandurin2018,Semenov2021,Muravev2007}.

The conductivity $\sigma$ of the 2D ES is considered in the framework of the dynamical Drude model without relaxation:
\begin{gather}
    \sigma_{xx}=\sigma_{yy}=\frac{n_s e^2}{m}\frac{i\omega}{\omega^2-\omega_c^2},\nonumber\\
    \sigma_{xy}=-\sigma_{yx}=\frac{n_s e^2}{m}\frac{\omega_c}{\omega^2-\omega_c^2},
    \label{eq:conductivity}
\end{gather}
where $e$, $m$, and $n_s$ are the charge, the effective mass, and the surface concentration of the electrons respectively, $\omega_c=|e|B/mc$ is the cyclotron frequency, and $B$ is the projection of the magnetic field onto the $z$-axis (therefore, the value of $\omega_c$ can be negative). The conductivity tensor is written in the ``clean'' limit, i.e., we consider the electron relaxation rate to be much smaller than the plasmon and cyclotron frequencies.

Using the local Ohm's law with the conductivity (\ref{eq:conductivity}), where the electric field is expressed through the charge density from Eq.(\ref{eq:potential}), we obtain the first equation, which describes the dynamics of the system. Together with the continuity equation we arrive at the following system of equations:
\begin{gather}
    -i\omega\bm{j}(x,y)+\omega_c \hat{R}\bm{j}(x,y)   
    =
    -v_{pl}^2\nabla_{2D}\delta\rho(x,y),\nonumber\\
    -i\omega\delta\rho(x,y)+\nabla_{2D}\cdot\bm{j}(x,y)=0.
    \label{eq:plasmons_main_equations}
\end{gather}
Here $\nabla_{2D}$ is the 2D Nabla operator, $\hat{R}$ is the 2D rotation matrix which rotates a vector clockwise by $\pi/2$ radians and $v_{pl}^2=4\pi d n_s e^2/\varepsilon m$.

Now, let us consider water waves in a shallow pool with depth $H$ on a planet that rotates with a frequency $\Omega$. The latitude of the pool is $\theta$, and geometric sizes are significantly smaller than the planet radius. We choose the Cartesian system whose $z$-axis is anti-directed along the gravitational acceleration $\bm{g}$. The frequency of the wave is $\omega$, and as before the perturbation of the water height $\delta h$ is proportional to $e^{-i\omega t}$. Then the linear dynamics of the rotational gravity shallow water waves is governed by the following system of equations \cite{Vallis2017}:
\begin{gather}
    -i\omega\bm{v}(x,y)+2\Omega\sin\theta\cdot\hat{R}\bm{v}(x,y)=-v_w^2\nabla_{2D}\frac{\delta h(x,y)}{H},\nonumber\\
    -i\omega\frac{\delta h(x,y)}{H}+\nabla_{2D}\cdot\bm{v}(x,y)=0,
    \label{eq:water_main_equations}
\end{gather}
where $v_w^2=gH$. Thus, in the considered approximations plasma waves in the fully screened limit, described by Eq.~(\ref{eq:plasmons_main_equations}), and shallow water waves, governed by Eq.~(\ref{eq:water_main_equations}), transform into each other after substitutions $\bm{j}/e n_s \leftrightarrow\bm{v}$, $\delta\rho/e n_s \leftrightarrow \delta h/H$, $\omega_c \leftrightarrow 2\Omega\sin\theta$, and $v_{pl} \leftrightarrow v_w$.

Finally, note that at the edge of the systems the normal component of the electrical current, or equivalently the hydrodynamic velocity, is zero. Thus, in the laterally confined systems under consideration, both the equations of motions and boundary conditions fully coincide. Therefore, all modes in a fully screened 2D ES and in a rotating shallow pool of the same shape will be completely identical. In particular, in a laterally infinite 2D ES magnetoplasmons have ``relativistic''-like dispersion on the wave vector $\omega_{pl}(\bm{q})=\sqrt{\omega_c^2+v_{pl}^2q^2}$. The same rotational gravity shallow water waves are well known as Poincare waves. For a half plane system there are also edge magnetoplasmons localized near the linear edge with the linear dispersion $\omega_{emp}(\bm{q})=v_{pl}q$, which exactly correspond to the Kelvin waves in hydrodynamics.

Below we analyze magnetoplasmons in a rectangular 2D ES and, according to the above, simultaneously find rotational gravity shallow water modes of a rectangular pool.

\section{\label{sec:rectangular_geometries} Magnetoplasmons in a rectangle}

\begin{figure}
    \includegraphics[width=\linewidth]{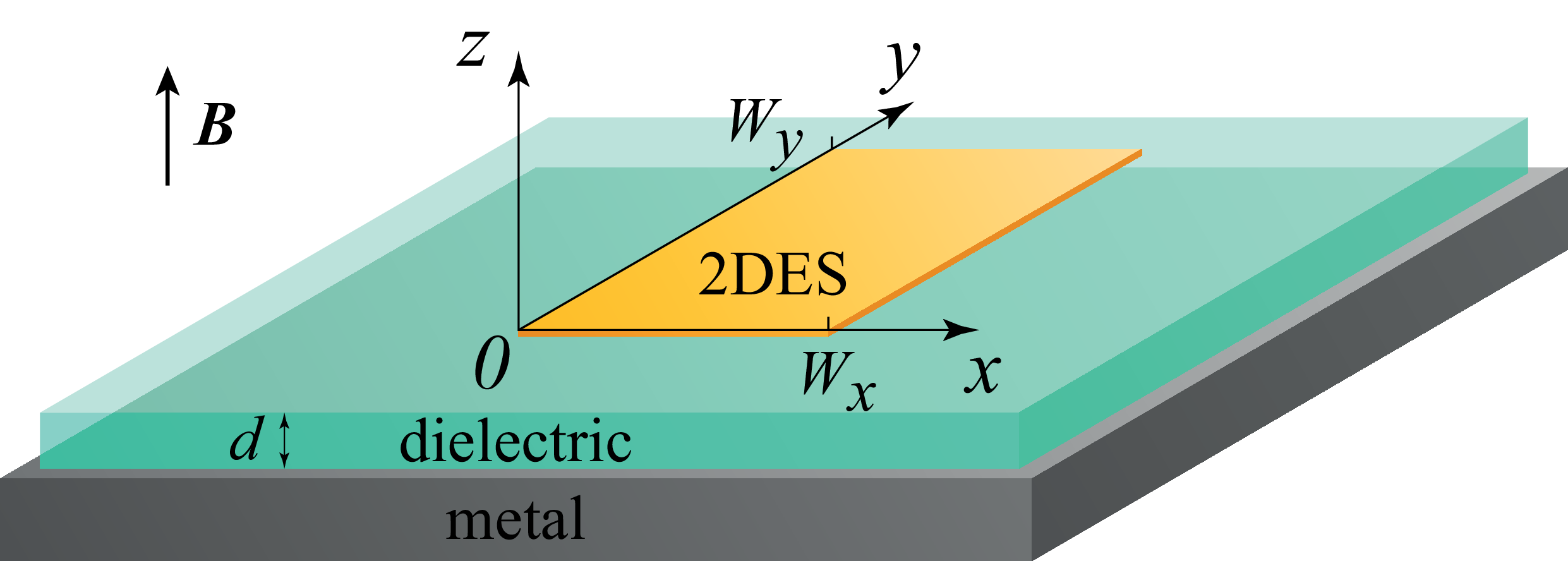}
    \caption{Schematic view of the system. A rectangle of a two-dimensional electron gas with sizes $W_x$ and $W_y$ is located at distance $d$ above an infinite metal gate. There is a substrate with the dielectric permittivity $\varepsilon$ between the gas and the metal. The system is placed in a perpendicular magnetic field.}
    \label{fig:scheme}
\end{figure}

Consider a rectangular 2D ES shown in Fig.\ref{fig:scheme}. Let the $x$ and $y$ axes be directed along the edges and the center of the coordinate system be placed in the rectangle corner. The transverse and longitudinal dimensions are equal to $W_x$ and $W_y$, respectively. From Eq.(\ref{eq:plasmons_main_equations}) we get
\begin{equation}\label{eq:current}
    \omega^2\bm{j}(x,y)+i\omega\omega_c\hat{R}\bm{j}(x,y)=-v_{pl}^2\hat{D}\bm{j}(x,y),
\end{equation}
where we introduce
\begin{equation}
    \hat{D}
    =
    \begin{pmatrix}
        \frac{\partial^2}{\partial x^2} & \frac{\partial^2}{\partial x \partial y}\\
        \frac{\partial^2}{\partial y \partial x} & \frac{\partial^2}{\partial x^2}
    \end{pmatrix},
    \quad
    \hat{R}
    =
    \begin{pmatrix}
        0 & 1\\
        -1 & 0
    \end{pmatrix}.
\end{equation}
Note that the operators $\hat{D}$ and $i\hat{R}$ are hermitian in the Hilbert space with the following scalar product
\begin{equation}
    \langle\bm{f}|\bm{g}\rangle = \int\limits_{0}^{W_x}\int\limits_{0}^{W_y}\bm{f}^*(x,y)\cdot\bm{g}(x,y)dxdy,
\end{equation}
where the normal components of the vector functions $\bm{f}$ and $\bm{g}$ to the edges of the considered system are equal to zero.

In the absence of a magnetic field, Eq. (\ref{eq:current}) turns into the eigenproblem of the operator $-v_{pl}^2\hat{D}$, which can be easily solved. The eigenvalues of the problem are the square of the plasmon frequencies
\begin{equation}\label{eq:frequency}
    \omega_{n_x,n_y}^2=v_{pl}^2\left[\left(\frac{n_x\pi}{W_x}\right)^2+\left(\frac{n_y\pi}{W_y}\right)^2\right],
\end{equation}
where $n_x$ and $n_y$ are pairs of positive integers except for $n_x=0$ and $n_y=0$ simultaneously. The integers $n_x$ and $n_y$ correspond to a number of half wavelengths fitting in the sizes $W_x$ and $W_y$, respectively. The current density is the eigenvector of the problem, which can be written using the ``bracket'' formalism in the following form:
\begin{gather}
    |n_x,n_y\rangle=
    A_{n_x,n_y}
    \begin{pmatrix}
        \frac{n_x}{W_x}\sin\left(\frac{n_x\pi}{W_x}x\right)\cos\left(\frac{n_y\pi}{W_y}y\right)\\
        \frac{n_y}{W_y}\cos\left(\frac{n_x\pi}{W_x}x\right)\sin\left(\frac{n_y\pi}{W_y}y\right)
    \end{pmatrix},\nonumber\\
    A_{n_x,n_y} = \frac{2}{\sqrt{\frac{n_x^2 W_y}{W_x}(1+\delta_{n_y,0})+\frac{n_y^2 W_x}{W_y}(1+\delta_{n_x,0})}},
    \label{eq:selfcurrent}
\end{gather}
The eigenvectors satisfy the normalization condition $\langle n_x,n_y|n_x',n_y'\rangle = \delta_{n_x,n_x'}\delta_{n_y,n_y'}$ where $\delta_{n,m}$ is the Kronecker delta.

To solve Eq. (\ref{eq:current}) in a magnetic field, we consider the magnetic term as a perturbation, which also depends on the frequency $\omega$, and expand the current density $\bm{j}$ in a complete set of functions being the unperturbed modes in the absence of a magnetic field
\begin{equation}\label{eq:expansion}
    |\bm{j}\rangle=\sum_{n_x,n_y}C_{n_x,n_y}|n_x,n_y\rangle.
\end{equation}
Then we substitute the expansion into Eq. (\ref{eq:current}) and consecutively multiply it by the unperturbed current (\ref{eq:selfcurrent}) to obtain the system of equations for the unknown coefficients $C_{n_x,n_y}$:
\begin{multline}\label{eq:system}
    \left(\omega^2-\omega_{n_x,n_y}^2\right) C_{n_x,n_y}+\\
    +\omega\omega_c \sum_{n_x',n_y'}C_{n_x',n_y'}\langle n_x,n_y|i\hat{R}|n_x',n_y'\rangle=0.
\end{multline}
After that, we cut the system (\ref{eq:system}), leaving enough coefficients corresponding the lowest frequencies. Thus, we obtain a finite set of algebraic equations, by solving which numerically or analytically, we find (approximately) the dependence of frequency on the magnetic field, i.e., magnetodispersion.

Before analyzing the results let us note that system (\ref{eq:system}) splits into two independent subsystems according to the parity of $n_x'+n_y'$. Indeed, the explicit expression for the matrix elements of the operator $i\hat{R}$ has the following form:
\begin{multline}\label{eq:Matrix_iR}
    \langle n_x,n_y|i\hat{R}|n_x',n_y'\rangle = \\\frac{4}{\pi^2}A_{n_x,n_y} A_{n_x',n_y'}\cdot
    \left(\frac{n_x^2}{n_x^2-{n_x'}^2}-\frac{n_y^2}{n_y^2-{n_y'}^2}\right)\cdot\\ \sin^2\left(\frac{\pi(n_x-n_x')}{2}\right) \sin^2\left(\frac{\pi(n_y-n_y')}{2}\right).
\end{multline}
If the parities of the numbers $n_x$ and $n_y$ are the same, i.e., $n_x+n_y$ is even, then the matrix elements (\ref{eq:Matrix_iR}) are nonzero only if the parities of the numbers $n_x'$ and $n_y'$ coincide. Correspondingly, for the different parities of the numbers $n_x$ and $n_y$, the matrix elements are nonzero if $n_x'+n_y'$ is odd. Therefore, system (\ref{eq:system}) divides into two independent subsystems: one of them involves the coefficients $C_{1,1}$, $C_{2,0}$, $C_{0,2}$, etc., the other $C_{0,1}$, $C_{1,0}$, $C_{1,2}$, $C_{2,1}$, etc.

\begin{figure}
    \includegraphics[width=0.9\linewidth]{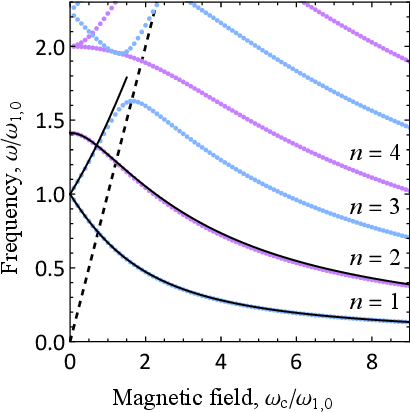}
    \caption{Dependence of the plasma frequency $\omega$ in a square on the cyclotron frequency divided by the frequency $\omega_{1,0}$ defined in Eq. (\ref{eq:frequency}). Dots correspond to the numerical calculations involving $80$ expansion coefficients in the system (\ref{eq:system}). Solid lines are the analytical result from Eqs. (\ref{eq:splited_01_10_modes}) and (\ref{eq:11_mode}).}
    \label{fig:square}
\end{figure}

Now let us discuss the results. At the beginning, we analyze the particular case of a square ($W_x=W_y\equiv W$). In Fig. \ref{fig:square} we plot the magnetodispersion of plasmon frequencies. Different colors indicate the magnetodispersions obtained from the two subsystems. In the absence of a magnetic field, modes with $n_x \neq n_y$ are degenerate at least twice since $\omega_{n_x,n_y} = \omega_{n_y,n_x}$. The magnetic field splits them. It is convenient to number the modes in order of their frequencies in strong magnetic fields. For example, modes $|0,1\rangle$ and $|1,0\rangle$ continuously transform into modes $n=1$ and $n=3$ in high magnetic fields. Leaving only two coefficients $C_{0,1}$ and $C_{1,0}$ in the system (\ref{eq:system}), we find their frequencies
\begin{equation}\label{eq:splited_01_10_modes}
    \omega_{n} = \sqrt{\left(\omega_{1,0}\right)^2+\left(0.41\omega_c\right)^2}+(n-2)0.41|\omega_c|,\quad n=1,3.
\end{equation}
The analytical result for the fundamental mode $n=1$, which is expected to be the edge magnetoplasmon in strong magnetic fields, is in amazing agreement for any magnetic field with the numerical result, which includes many more coefficients. This mode has negative magnetodispersion. At the same time, the frequency of mode $n=3$ at first increases with increasing magnetic field and tends to cross other modes. However, as in \cite{Zarezin2023}, there is avoided crossing between this mode and the next one with a higher frequency marked in blue. Therefore, for a correct description of this mode at large magnetic fields, it is necessary to take into account modes with higher $n_x$, $n_y$, and expression (\ref{eq:splited_01_10_modes}) is valid only for small magnetic fields. The behavior of these two modes is shown in Fig.\ref{fig:square} by a solid line for frequencies below $1.8\omega_{1,0}$.

To obtain an approximate magnetodispersion for mode $n=2$, we leave coefficients $C_{1,1}$, $C_{0,2}$, and $C_{2,0}$ in Eq.(\ref{eq:system}). After some algebra we derive the following expression:
\begin{equation}\label{eq:11_mode}
    \omega_2 = \omega_{1,1}\sqrt{1.5+a^2-\sqrt{0.25+3a^2+a^4}},
\end{equation}
where $a=0.54\omega_c/\omega_{1,1}$. This magnetodispersion is also shown in Fig. \ref{fig:square} by a solid line and demonstrates excellent accuracy for all magnetic fields.

\begin{figure*}
    \includegraphics[width=1\linewidth]{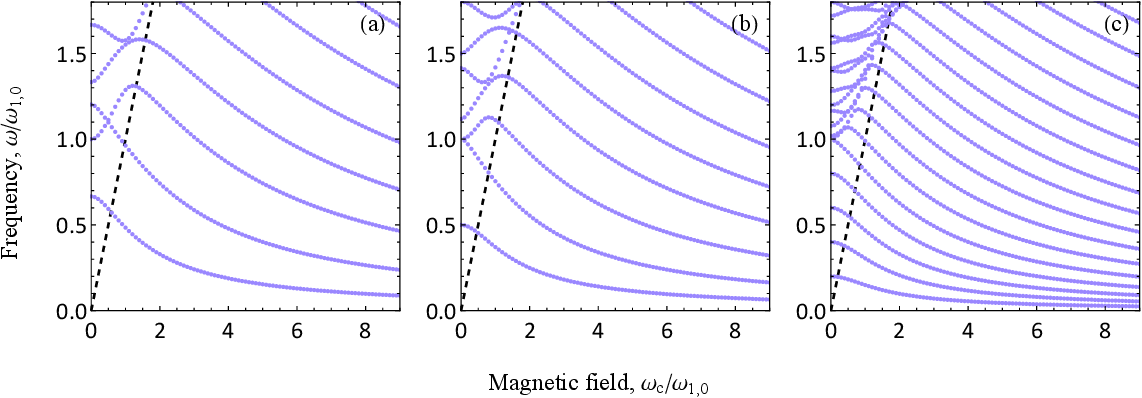}
    \caption{Dependence of the plasma frequency $\omega$ in a rectangle on the cyclotron frequency divided by the frequency $\omega_{1,0}$ defined in Eq. (\ref{eq:frequency}). Results are presented for rectangles with different sizes: (a) $W_y=1.5W_x$, (b) $W_y=2W_x$, and (c) $W_y=5W_x$. Dots correspond to the numerical calculations.}
    \label{fig:rectangular}
\end{figure*}

Now, let us discuss plasma modes in a rectangle with an arbitrary ratio of width and length. A number of magnetodispersions are shown in Fig. \ref{fig:rectangular} for different aspect ratios. Now, the lowest mode is nondegenerate if $W_x \neq W_y$. The first order correction in the magnetic field vanishes for nondegenerate modes since it is proportional to the average value of the operator $i\hat{R}$ which is always equal to zero. Details of these calculations are presented in Appendix \ref{app:B}. Therefore, in weak magnetic fields, the nondegenerate modes behave quadratically in the magnetic field. In strong magnetic fields, the frequency generally decreases as $1/\omega_c$. Thus, for the modes already discussed for a square, this is clearly visible from Eqs. (\ref{eq:splited_01_10_modes}) and (\ref{eq:11_mode}). In other cases, one can expect that coefficients $C_{n_x,n_y}$ do not diverge in $\omega_c$, and therefore, to compensate the linear divergence in $\omega_c$ in the second term of Eq. (\ref{eq:system}), the frequency $\omega$ should vanish as $\alpha\,\omega_{1,0}^2/\omega_c$, where $\alpha$ is a constant. To estimate this constant, we plug $\omega=\alpha\,\omega_{1,0}^2/\omega_c$ into Eq. (\ref{eq:system}), neglect the frequency $\omega$ compared to $\omega_{n_x,n_y}$, and find that the coefficients $1/\alpha$ are the eigenvalues of the matrix $\langle n_x,n_y|i\hat{R}|n_x',n_y'\rangle/\omega_{n_x,n_y}^2$, which can be evaluated numerically. For example, $\alpha_1\approx 1.2$, $\alpha_2\approx 3.6$, and $\alpha_3\approx 7.0$ for modes with $n=1,2,3$. This asymptotic behavior in a high magnetic field is the important feature of the rectangular shape. Thus, within the same approximations in a disk, half plane, or stripe, the frequency of the lowest modes in high magnetic fields tends to a nonzero value \cite{Volkov1988,Zagorodnev2023}.

\section{\label{sec:conclustion} Discussion and conclusion}

We have examined magnetoplasmons in a rectangular two-dimensional ES in the fully screened limit neglecting non-linearity and at the same time have found linear rotational gravity shallow water waves of the same geometry. Although we have pointed to the exact correspondence between these two systems, it is worth noting the difference between them. First, modes of the systems exist in different frequency ranges. For example, in modern high mobility 2D ESs, such as quantum wells or graphene, the typical frequencies for plasmons lie in the sub and terahertz range. As for waves in shallow water, affected by gravity and Earth's rotation, we deal with the sub and millihertz frequency ranges. For example, on the Earth's surface near the geographic poles, the Coriolis frequency takes the maximum value of 23 $\mu$Hz. Second, Eq. (\ref{eq:potential}) was derived in the quasistatic approximation, when electromagnetic retardation effects and, in particular, the retardation damping of plasmons are absent. However, when taking into account the effects of electromagnetic retardation (i.e., in the next order in the speed of light), radiative damping will arise, which in principle does not have an analogy for waves in hydrodynamics. Third, we have analyzed both the plasmons and the shallow water waves only in the linear regime. If the analogy extends to the nonlinear regime, it is possible to give an example of phenomena that are absent in electrodynamics. For instance, there is the hydrodynamic wave breaking, which corresponds to the case when the wave crest hangs over the water surface. It means that the height of the water is multivalued at the points of the crest. In electrodynamics it is excluded because a multivalued electron density would lead to a nonunique value of the electrical potential at a space point.

Let us emphasize that there was no metal gate in the experimental study \cite{Zarezin2023}, while we have considered the fully screened limit. In our case the gate significantly affects the plasma frequencies, therefore, comparison of our calculations with the experimental data can be made only qualitatively. However, let us point out that, despite the lack of experimental data in the limit, this regime has several distinctive advantages. First, it can be analyzed analytically with controllable accuracy. An analytical description of ungated rectangular systems is a challenging problem. In particular, there are no exact solutions (the natural frequencies, charge, and current distributions) even in a zero magnetic field. Whereas in the fully screened limit without a magnetic field one can find the exact solutions for all modes. It is this feature that paves the way to an analytical description of magnetodispersion with a given accuracy. Second, the near gate screens the electrical fields of the 2D electron system, preventing the interaction with other metal parts of the system, such as gate for depletion, waveguide, etc., which are always present in the system and can affect plasma resonances. Thus, in our opinion, a fully screened 2D ES is more suitable for comparing calculations with experimental data.

\section{\label{sec:acknowledgments} Acknowledgments}
The authors are grateful to Andrey Zabolotnykh for valuable discussions. This work was carried out within the framework of the state task of the Kotelnikov Institute of Radioengineering and Electronics of the Russian Academy of Sciences. D.A.R. especially thanks the Foundation for the Advancement of Theoretical Physics and Mathematics ``BASIS'' (project No. 21-1-5-133-1).

\appendix
\section{\label{app:A} Local capacitance in the fully screened limit}
The Poisson equation for the electric potential caused by a perturbation of the charge density is
\begin{equation}\label{eq:Poison}
    \nabla\cdot\varepsilon(z)\nabla\varphi(x,y,z)=-4\pi\delta\rho(x,y)\delta(z),
\end{equation}
where $\nabla$ is the Nabla operator, $\delta(z)$ is the Dirac function and the dielectric permittivity $\varepsilon(z)$ such that $\varepsilon(z>0)=\kappa$ and $\varepsilon(-d<z<0)=\varepsilon$. Applying the Fourier transform to $x$ and $y$ coordinates, we get the differential equation
\begin{equation}\label{eq:fourier_potential}
    \left[\frac{\partial}{\partial z}\varepsilon(z)\frac{\partial}{\partial z}-\varepsilon(z)q^2\right]\phi(\bm{q},z)=-4\pi\delta\rho(\bm{q})\delta(z),
\end{equation}
where $\bm{q}=(q_x,q_y)^T$ is the parameter of the transform, $q=\sqrt{q_x^2+q_y^2}$ is an amplitude of the $\bm{q}$ parameter, $\varphi(\bm{q},z)$ and $\delta\rho(\bm{q},z)$ are the Fourier images of the potential and the charge density perturbation, respectively. We look for a continuous solution, which is equal to zero in the plane $z=-d$ and vanishes at $z\rightarrow\infty$. In this case, we can write that 
\begin{equation}
    \begin{cases}
        \phi(\bm{q},z>0)=\Phi(\bm{q}) e^{-q z}\sinh q d,\\
        \phi(\bm{q},z<0)=\Phi(\bm{q})\sinh q(z+d),
    \end{cases}
\end{equation}
where $\Phi(\bm{q})$ is a coefficient. To determine the coefficient $\Phi(\bm{q})$ we use the condition
\begin{equation}
    \kappa\frac{\partial}{\partial z}\phi(\bm{q},z)\Big|_{z=+0}-\varepsilon\frac{\partial}{\partial z}\phi(\bm{q},z)\Big|_{z=-0}=-4\pi\delta\rho(\bm{q})
\end{equation}
dictated by the presence of the Dirac function in the right side of Eq.(\ref{eq:fourier_potential}). Thus, we obtain a connection between the Fourier images of the potential and the charge density perturbation:
\begin{equation}
    \varphi(\bm{q},z)=2\pi\rho(\bm{q})\frac{e^{-q(|z|-d)}-e^{-q(z+d)}}{q\left(\kappa\sinh qd+\varepsilon\cosh qd\right)}.
\end{equation}
Finally, applying the inverse Fourier transform and considering the 2D ES plane $z=0$, we have the integral connection
\begin{equation}
    \varphi(x,y,z=0)=4\pi\int\limits_{-\infty}^{\infty}\int\limits_{-\infty}^{\infty} G(x-x',y-y')\rho(x',y')dx'dy',
\end{equation}
where the integral kernel is given by the expression
\begin{equation}
    G(x,y)=\int\limits_{-\infty}^{\infty}\int\limits_{-\infty}^{\infty}\frac{e^{iq_x x+iq_y y}}{q\left(\kappa+\varepsilon\coth qd\right)}\frac{dq_x dq_y}{(2\pi)^2} .
\end{equation}
Expanding the term $\left(\kappa+\varepsilon\coth qd\right)^{-1}$ as $qd/\varepsilon$ at the condition $qd\ll 1$, the kernel becomes the two-dimensional Dirac delta function and, therefore, we obtain Eq.(\ref{eq:potential}).

\section{\label{app:B} Perturbation theory for nondegenerate modes}
In this part we find the first nonzero contribution of the magnetic field to the frequency of nondegenerate modes. Let us consider system (\ref{eq:expansion}). To reduce expressions, we denote the set of quantum numbers $n_x$ and $n_y$ by one number $\bm{n}$. We find the coefficients and the frequency in the following form:
\begin{gather}
    C_{\bm{n}}(\bm{N})=\delta_{\bm{n},\bm{N}} + C_{\bm{n}}^{(1)}(\bm{N})\omega_c + C_{\bm{n}}^{(2)}(\bm{N})\omega_c^2 + ...\\
    \omega^2(\bm{N})=\omega_{\bm{N}}^{2} + \left(\omega^{(1)}(\bm{N})\right)^2\omega_c + \left(\omega^{(2)}(\bm{N})\right)^2\omega_c^2 + ...,
\end{gather}
where $\delta_{\bm{n},\bm{N}}$ is the two-dimensional Kronecker delta and $\bm{N}$ corresponds to the set of numbers of the considered nondegenerate mode. Substituting this representation into Eq.(\ref{eq:expansion}) and collecting terms of the similar order of the magnetic field, we consistently get equations for the perturbation of the frequency and coefficients. In the first order we have the equation
\begin{multline}
    \omega_{\bm{N}}\langle \bm{n}|i\hat{R}|\bm{N}\rangle + \left(\omega^{(1)}(\bm{N})\right)^2\delta_{\bm{n},\bm{N}} - \\
    - \left(\omega_{\bm{n}}^2-\omega_{\bm{N}}^2\right) C_{\bm{n}}^{(1)}(\bm{N})=0,
\end{multline}
which leads to the following result:
\begin{align}
    \left(\omega^{(1)}(\bm{N})\right)^2=-\omega_{\bm{N}}\langle \bm{N}|i\hat{R}|\bm{N}\rangle=0,\\
    C_{\bm{n}}^{(1)}(\bm{N})=\frac{\omega_{\bm{N}}\langle \bm{n}|i\hat{R}|\bm{N}\rangle}{\omega_{\bm{n}}^2-\omega_{\bm{N}}^2},\quad \bm{n}\neq \bm{N}.
\end{align}
Taking into account that $\omega^{(1)}(\bm{N})=0$, in the second order we obtain the following equation:
\begin{multline}
    \omega_{\bm{N}}\sum\limits_{\bm{n}'}C_{\bm{n}'}^{(1)}\langle \bm{n}|i\hat{R}|\bm{n}'\rangle + \left(\omega^{(2)}(\bm{N})\right)^2\delta_{\bm{n},\bm{N}} - \\
    - \left(\omega_{\bm{n}}^2-\omega_{\bm{N}}^2\right) C_{\bm{n}}^{(2)}(\bm{N})=0.
\end{multline}
and, consequently,
\begin{gather}
    \left(\omega^{(2)}(\bm{N})\right)^2=-\omega_{\bm{N}}\sum\limits_{\bm{n}'}C_{\bm{n}'}^{(1)}\langle \bm{N}|i\hat{R}|\bm{n}'\rangle,
    \\
    C_{\bm{n}}^{(2)}(\bm{N})=\frac{\omega_{\bm{N}}}{\omega_{\bm{n}}^2-\omega_{\bm{N}}^2}\sum\limits_{\bm{n}'}C_{\bm{n}'}^{(1)}\langle \bm{N}|i\hat{R}|\bm{n}'\rangle,\quad \bm{n}\neq \bm{N}.
\end{gather}
The undefined coefficients $C_{\bm{N}}^{(1)}$ and $C_{\bm{N}}^{(2)}$ are found from the normalization condition:
\begin{multline}
    1 = \sum\limits_{\bm{n}}|C_{\bm{n}}(\bm{N})|^2=1 + \left(C_{\bm{N}}^{(1)}(\bm{N})+C_{\bm{N}}^{(1)*}(\bm{N})\right)\omega_c + \\
    + \left(C_{\bm{N}}^{(2)}(\bm{N}) + C_{\bm{N}}^{(2)*}(\bm{N}) + \sum\limits_{\bm{n}} \big|C_{\bm{n}}^{(1)}(\bm{N})\big|^2\right)\omega_c^2 + ....
\end{multline}
Therefore, we get
\begin{equation}
    C_{\bm{N}}^{(1)}(\bm{N})=0,\quad C_{\bm{N}}^{(2)}(\bm{N})=-\frac{1}{2}\sum\limits_{\bm{n}} \big|C_{\bm{n}}^{(1)}(\bm{N})\big|^2.
\end{equation}
Thus, the frequency of the nondegenerate mode is given by
\begin{multline}
    \omega(\bm{N}) = \omega_{\bm{N}} + \frac{\left(\omega^{(2)}(\bm{N})\right)^2}{2\omega_{\bm{N}}}\omega_c^2 + ... = \\
    = \omega_{\bm{N}}\left[1-\frac{1}{2}\sum\limits_{\bm{n}'\neq \bm{N}}\frac{\Big| \langle \bm{n}'|i\hat{R}|\bm{N}\rangle \Big|^2}{\omega_{\bm{n}'}^2-\omega_{\bm{N}}^2}\omega_c^2 + ...\right].
\end{multline}

\bibliography{bibl}

\end{document}